\documentclass[11pt]{article}

\usepackage{amsmath,amssymb}

\newcommand{\Slash}[1]{\ooalign{\hfil/\hfil\crcr$#1$}}
\newcommand{\Td}[2]{\frac{d{#1}}{d{#2}}}

\newcommand{\del}{\partial}
\newcommand{\half}{\frac{1}{2}}
\newcommand{\mbf}{\boldsymbol}
\newcommand{\ls}{\ \ \ \ \ }
\newcommand{\ol}{\overline}
\newcommand{\dps}{\displaystyle}
\newcommand{\kahler}{K\"{a}hler }

\newcommand{\bsubeq}{\begin{subequations}}
\newcommand{\esubeq}{\end{subequations}}
\newcommand{\vs}[1]{\vspace{#1 mm}}

\begin{document}

\allowdisplaybreaks{
\setcounter{page}{0}

\begin{titlepage}

{\normalsize
\begin{flushright}
OU-HET 382\\
hep-th/0104184\\
April 2001
\end{flushright}
}
\bigskip

\begin{center}
{\LARGE\bf Supersymmetric Nonlinear Sigma Models \\

on Ricci-flat \kahler Manifolds with $\mbf{O(N)}$ Symmetry}

\vs{10}

\bigskip
{\renewcommand{\thefootnote}{\fnsymbol{footnote}}
{\Large\bf Kiyoshi Higashijima\footnote{
     E-mail: {\tt higashij@phys.sci.osaka-u.ac.jp}},
 Tetsuji Kimura\footnote{
     E-mail: {\tt t-kimura@het.phys.sci.osaka-u.ac.jp}} {\large and}
 Muneto Nitta\footnote{
     E-mail: {\tt nitta@het.phys.sci.osaka-u.ac.jp}}
}}

\setcounter{footnote}{0}
\bigskip

{\large\sl
Department of Physics,
Graduate School of Science, Osaka University, \\
Toyonaka, Osaka 560-0043, Japan \\
}
\end{center}
\bigskip


\begin{abstract}
We propose a class of ${\cal N}=2$ supersymmetric nonlinear sigma 
models on the Ricci-flat \kahler manifolds with $O(N)$ symmetry.
\end{abstract}

\end{titlepage}


String theory propagating in a curved spacetime is described 
by a conformally invariant nonlinear sigma model in two dimensions.  
Spacetime supersymmetry of the string theory requires ${\cal N}=2$ 
worldsheet supersymmetry. The conformal invariance is realized 
in finite field theories with vanishing $\beta$-functions. 
In this letter, we propose a class of nonlinear sigma models 
on Ricci-flat \kahler manifolds with $O(N)$ symmetry.


Two-dimensional ${\cal N}=2$ supersymmetric nonlinear sigma models 
are described by (anti-)chiral superfields:
$\phi =  A + \sqrt{2} \theta \psi + \theta \theta F$ 
($\phi^{\dagger} =  A^* + \sqrt{2} \; \bar{\theta} \bar{\psi} 
+ \bar{\theta} \bar{\theta} F^*$)
where $A$, $\psi$ and $F$ are 
complex scalar, Dirac fermion, auxiliary scalar fields, respectively.  
To define supersymmetric nonlinear sigma models on 
the Ricci-flat \kahler manifolds with $O(N)$ symmetry, 
we prepare dynamical chiral superfields $\phi^i$ 
($i=1,\cdots,N$; $N \geq 2$), 
belonging to the vector representation of $O(N)$, 
and an auxiliary chiral superfield $\phi_0$, 
belonging to an $O(N)$ singlet. 
The most general Lagrangian with $O(N)$ symmetry, 
composed of these chiral superfields, is given by 
\begin{align}
{\cal L} \ = \ 
\int \! d^4 \theta \; {\cal K} (x) 
+ \int \! d^2 \theta \; \phi_0 \left( \sum_{i=1}^N \phi^i \phi^i
- a^2 \right) 
+ \int \! d^2 \ol{\theta} \; \phi_0^{\dagger} \left( \sum_{i=1}^N
\phi^{\dagger}{}^i \phi^{\dagger}{}^i - a^2 \right) \; , \label{most}
\end{align}
where $x$ is the $O(N)$ invariant
defined by
\begin{align}
 x \ = \ \sum_{i=1}^N \phi^{\dagger}{}^i \phi^i \; ,
\end{align}
and ${\cal K} (x)$ is an arbitrary function of $x$. 
In the Lagrangian (\ref{most}), 
we can assume that $a^2$ is
a positive real constant, using the field redefinition. 
By the integration over the auxiliary field $\phi_0$,  
we obtain the constraint among the superfields $\phi^i$, 
$\sum_{i=1}^N \phi^i \phi^i = a^2$, 
whose bosonic part is
\begin{align}
 \sum_{i=1}^N A^i A^i \ &= \ a^2 \; . \label{constraint}
\end{align}
The manifold defined by this constraint 
with the \kahler potential ${\cal K}$ 
is a {\it non-compact} 
\kahler manifold with the complex dimension $N-1$, 
where the $O(N)$ symmetry acts as a holomorphic isometry. 
One of the component, say $A^N$, can be expressed in terms of the
independent fields $A^a$ ($a=1,\cdots,N-1$) 
\begin{align}
 A^N \ &= \ \sqrt {a^2 - \sum_{a=1}^{N-1} A^a A^a} \; . \label{AN}  
\end{align}

To obtain the Ricci-flat \kahler manifold with $O(N)$ symmetry,
we calculate
the Ricci form and solve the Ricci-flat condition. 
To do this, we first calculate the \kahler metric $g_{a b^*} = \del_a
\del_{b^*} {\cal K}$ of the manifold:
\begin{align}
 g_{a b^*} 
 \ &= \ {\cal K}' \Big( \delta_{a b} 
     + \frac{A^a A^{*b}}{A^{*N} A^N} \Big) 
 + {\cal K}'' \Big( A^{*a} - \frac{A^{*N}}{A^N} A^a \Big) 
              \Big( A^b - \frac{A^N}{A^{*N}} A^{*b} \Big) \; , 
 \label{metric-N}
\end{align}
where the prime denotes differentiation with respect to $x$, 
and $A^N$ is given by (\ref{AN}). 
Here we have used the same letter $x$ 
for its lowest component: $x= \sum_{i=1}^N A^{*i}A^i$. 
Using this metric the nonlinear sigma model Lagrangian is written as
\begin{align}
{\cal L} \ &= \ g_{a b^*} \del_{\mu} A^a \del^{\mu} A^{*b} + i g_{a
b^*} \ol{\psi}^b ( \Slash{D} \psi)^a + \frac{1}{4} R_{a b^*
c d^*} \psi^a \psi^c \ol{\psi}^b \ol{\psi}^d \; ,
\end{align}
where $R_{a b^* c d^*}$ is the Riemann curvature tensor and $(D_{\mu}
\psi)^a = \del_{\mu} \psi^a + \del_{\mu} A^b \Gamma^a{}_{b c}
\psi^c$~\cite{Zu}.
Without loss of generality, 
an arbitrary point on the manifold can be 
transformed by the $O(N)$ symmetry to 
\begin{align}
 A^{\alpha} \ &= \ 0 \ \ (\alpha = 1 , 2 , \cdots , N-2) \; , \ \ \
 A^{N-1} \ = \ z \; , \ \ \ 
 A^N \ = \ \sqrt{a^2 -z^2} \; , \label{vev-N}
\end{align}
where $z$ is complex. 
The \kahler metric on this point is  
\begin{align}
  g_{a b^*} \ &= \ 
\left(\begin{array}{cc}
  {\cal K}' \delta_{\alpha \beta^*} & 0 \\
   0 & \dps \frac{1}{A^{*N} A^N} 
       \big[ {\cal K}' \cdot x + {\cal K}'' \cdot ( x^2 - a^4) \big]  
\end{array} \right) \; , \label{gen.metric}
\end{align}
and its determinant is given by 
\begin{align}
\det g_{a b^*} 
 \ &= \ \frac{1}{A^{*N} A^N }
  \left[ ({\cal K}')^{N-1} \cdot x + ({\cal K}')^{N-2} ({\cal
      K}'') \cdot (x^2 - a^4) \right] \; . \label{det-N}
\end{align}

The Ricci form is given by 
$R_{a b^*} = - \del_a \del_{b^*} \log \det g$. 
Hence, the Ricci-flat condition $R_{a b^*}=0$ 
is equivalent to the condition that 
the determinant is a constant up to
products of holomorphic and anti-holomorphic functions:
$\det g_{a b^*} \equiv \mbox{(constant)} \times F (z) \times
F^* (z^*)$. From 
this condition, we obtain a differential equation
\begin{align}
({\cal K}')^{N-1} \cdot x 
 + ({\cal K}')^{N-2} ({\cal K}'') \cdot (x^2- a^4) \ &= \ c \; , 
 \label{ricci-flat-cond}
\end{align}
where $c$ is a constant. 
Using this equation,  
the metric (\ref{metric-N}) can be rewritten as 
\begin{align}
g_{a b^*} \ &= \ \frac{{\cal K}'}{A^{*N} A^N} \Big[ \delta_{a b}
A^{*N} A^N + A^a A^{*b} + \frac{c \cdot ({\cal
K}')^{-(N-1)} - x}{x^2 - a^4} \big( A^{*a} A^N - A^a A^{*N} \big)
\big( A^b A^{*N} - A^{*b} A^N \big) \Big] \; , \label{ricci-flat-metric}
\end{align}
where $A^N$ is given by (\ref{AN}). 
Therefore, we only need the solution of ${\cal K}'$ 
but not ${\cal K}$ itself,  
to calculate the Ricci-flat \kahler metric.

To solve
the nonlinear 
differential equation (\ref{ricci-flat-cond}), 
we transform it to a linear differential equation:
\begin{align}
\Td{y(x)}{x} + (N-1) \frac{x}{x^2 - a^4} y (x) \ &= \ \frac{c
(N-1)}{x^2 - a^4} \; , \ \ \ y (x) \ = \ ({\cal K}')^{N-1} \; .
\end{align}
The general solution of $y(x)$ is immediately obtained as  
\begin{align}
 y(x) \ &= \ 
 \Big( \frac{1}{x^2 - a^4} \Big)^{\frac{N-1}{2}} 
 \Big\{ y_0 (x_0^2 - a^4)^{\frac{N-1}{2}} 
      + c (N-1) \int_{x_0}^x \! d t \, (t^2 - a^4)^{\frac{N-3}{2}}
 \Big\} \; ,
\end{align}
where $x_0$ is an integration constant and $y_0 = y(x_0)$. 
In order to obtain the finite solution at $x = a^2$, 
we must set the parameter $x_0$ the solution of  
\begin{align}
y_0 (x_0^2 - a^4)^{\frac{N-1}{2}} + c (N-1) \int_{x_0}^{a^2} \! d t
(t^2 - a^4)^{\frac{N-3}{2}} \ = \ 0 \; .
\end{align}
Using this equation, 
we obtain an integral representation of
$y(x)$:
\begin{align}
&y(x) \ = \ ({\cal K}')^{N-1} \ = \ c (N-1) \Big( \frac{1}{x^2 - a^4}
\Big)^{\frac{N-1}{2}} \int_{a^2}^x \! d t \, (t^2 -
a^4)^{\frac{N-3}{2}} \; . \label{sol-N}
\end{align}
By performing the integration 
we can express $y(x)$ 
using the hypergeometric function $F(\alpha, \beta, \gamma ; z)$: 
\begin{align}
({\cal K}')^{N-1}
\ &= \ \frac{c (N-1)}{2 \Gamma (N/2)} (-a^4)^{\frac{N-3}{2}} 
 \Big(
 \frac{1}{x^2 - a^4} \Big)^{\frac{N-1}{2}} \nonumber \\
& \ls \times 
 \Big[ 2 x \Gamma \left(\frac{N}{2} \right) \cdot 
 F \left( \half , - \frac{N-3}{2} , \frac{3}{2} ; 
         \frac{x^2}{a^4} \right)
  - a^2 \sqrt{\pi} 
 \Gamma \left( \frac{N-1}{2} \right) \Big] \; . \label{hyper-N}
\end{align}
We can also obtain the Ricci-flat metric by substituting 
${\cal K}'$ to (\ref{ricci-flat-metric}).
We thus have obtained $(N-1)$-dimensional 
Ricci-flat \kahler manifolds with $O(N)$ symmetry. 

\vs{5}

When $N$ is odd ($N = 2m + 1$ and $m = 1, 2, \cdots$), 
the hypergeometric function reduces to a polynomial:
\begin{align}
({\cal K}')^{2m} \ &= \ 2 c \sum_{n=0}^{m-1} {}_{m-1} C_n \frac{m}{2n+1}
(-a^4)^{m-n-1} \Big[ \frac{x^{2n+1} - (a^2)^{2n+1}}{(x^2 - a^4)^m}
\Big] \; , \ \ {}_{m} C_n \ = \ \frac{m!}{n! (m-n)!} \; .
\end{align}
If $N$ is even ($N=2m+2$ and $m = 1,2,\cdots$) except for $N=2$ [see
(\ref{sol-2}), below, for the $N=2$ solution], 
the solution can be written as
\begin{align}
({\cal K}')^{2m+1} \ &= \ \frac{c (2m + 1)}{a^2 (\tan
\theta_x)^{2m+1}} \sum_{n=0}^m {}_m C_n (-1)^{m-n}
\nonumber \\
& \ \ \ \ \times \frac{(2 n - 1)!!}{(2n)!!} \Big\{ \frac{\sin
\theta_x}{(\cos \theta_x)^{2 n}} \sum_{r=0}^{n-1} \frac{(2n -
2r-2)!!}{(2n - 2r - 1)!!} (\cos \theta_x)^{2r} + \log \Big| \tan \big
( \frac{\theta_x}{2} + \frac{\pi}{4} \big) \Big| \Big\} \; ,
\end{align}
where $\theta_x = \arccos (a^2/x)$. 
Explicit expressions of 
${\cal K}'$ for $N=2$ to $7$ are 
\bsubeq
\begin{align}
N \ = \ 2 \ &: \ 
  {\cal K}' \ = \ 
  \frac{c}{\sqrt{x^2 - a^4}} \log \Big(
  \frac{x + \sqrt{x^2 - a^4}}{a^2} \Big) \; , \label{sol-2} \\
N \ = \ 3 \ &: \ 
  ({\cal K}')^2  \ = \ \frac{2 c}{x+a^2} \; , \label{sol-3} \\
N \ = \ 4 \ &: \ 
  ({\cal K}')^3 \ = \ \frac{3 c}{2} \frac{1}{(x^2 - a^4)^{3/2}} 
  \Big[ x \sqrt{x^2 - a^4} - a^4 \log \Big( \frac{x +
  \sqrt{x^2 - a^4}}{a^2} \Big) \Big] \; , \label{sol-4} \\
N \ = \ 5 \ &: \ 
  ({\cal K}')^4 \ = \ \frac{4 c}{3} \frac{x + 2 a^2}{(x + a^2)^2} \; ,\\
N \ = \ 6 \ &: \ 
  ({\cal K}')^5 \ = \ - \frac{5 c}{8} \frac{1}{(x^2 - a^4)^{5/2}} 
  \Big[ ( -2 x^3 + 5 a^4 x ) \sqrt{x^2 - a^4} 
    - 3 a^8 \log \Big( \frac{x + \sqrt{x^2 - a^4}}{a^2} 
  \Big) \Big] \; , \\
N \ = \ 7 \ &: 
  \ ({\cal K}')^6 \ = \ \frac{2 c}{5} 
  \frac{3 x^2 + 9 a^2 x + 8 a^4}{(x + a^2)^3} \; .
\end{align}
\esubeq

For definiteness, we give the explicit solution of ${\cal K}$ 
for $N=2$ and $N=3$. 
For $N=2$, ${\cal K}$ can be obtained as
\begin{align}
 {\cal K}_{N=2} \ &= \ 
 \frac{c}{2} \Big[ \log \Big( \frac{x + \sqrt{x^2 - a^4}}{a^2}\Big) 
             \Big]^2 \; . 
\end{align}
By the field redefinition 
$A^1 = a \cos \varphi$ ($\varphi$ $\in$ {\bf C}), 
this \kahler potential becomes 
\begin{align}
{\cal K}_{N=2} \ &= \ c \varphi^* \varphi + F (\varphi) + F^*
(\varphi^*) \; .
\end{align} 
Here, $F(\varphi)$ and $F^* (\varphi^*)$ are holomorphic and
anti-holomorphic functions, respectively, 
which can be eliminated by the \kahler transformation. 
Thus we obtain the free field theory with the flat metric 
$g_{\varphi \varphi^*}=c$ as we expected,
since the Ricci-flatness implies the vanishing Riemann curvature in
real two-dimensional manifolds. 
The solution of ${\cal K}$ for $N=3$ 
can be calculated, to yield
\begin{align}
 {\cal K}_{N=3} \ = \ \sqrt{8c} \sqrt{x+a^2} \; .
\end{align}
The Ricci-flat \kahler metric for $N=3$ is 
\begin{align}
g_{a b^*} \ &= \ \frac{\sqrt{2 c}}{(x+a^2)^{3/2}} 
 \left[ (x+a^2)
 \Big( \delta_{a b} + \frac{A^a A^{*b}}{A^{*3} A^3} \Big) - \half \Big(
 A^{*a} - \frac{A^{*3}}{A^3} A^a \Big) \Big( A^b - \frac{A^3}{A^{*3}}
 A^{*b} \Big) \right] \; ,
\end{align}
where $A^3$ is given in (\ref{AN}). 
This defines a (real) four-dimensional hyper-\kahler manifold 
with $O(3)$ symmetry, the Eguchi-Hanson space~\cite{Eguchi}. 
The metric for $N=4$ calculated from (\ref{sol-4}) coincides with 
the metric of the deformed conifold, 
obtained earlier in~\cite{Candelas}.

\vs{5}

We discuss the limit 
of vanishing $a^2$. 
In this limit, the manifold becomes a conifold, 
in which the point represented by $z=0$ in (\ref{vev-N}) is singular.
In this limit, we can obtain the explicit solution of 
${\cal K}$ for any $N(> 2)$, given by
\begin{align}
{\cal K} \ &= \ \Big( c \frac{N-1}{N-2} \Big)^{\frac{1}{N-1}}
\int_{x_0}^x \! d t \, t^{\frac{-1}{N-1}} \ = \ \frac{1}{c} \Big( c
\frac{N-1}{N-2} \Big)^{\frac{N}{N-1}} \cdot x^{\frac{N-2}{N-1}}
+ \mbox{(constant)} \;
. \label{a-zero}
\end{align}
In the $N \to \infty$ limit, 
the \kahler potential (\ref{a-zero}) becomes the simplest form:
\begin{align}
\lim_{N \to \infty} {\cal K} \ &= \ x \; . 
\end{align}
Therefore, when we discuss non-perturbative effects
of the sigma model on the conifold 
defined by (\ref{a-zero}), 
using the $1/N$ expansion method, 
it is sufficient to consider the simplest \kahler potential
${\cal K} = x$ instead of (\ref{a-zero}).


If we prepare invariants of other groups 
in the superpotential of (\ref{most}), 
we would obtain Ricci-flat \kahler manifolds 
with other symmetry. 

We have studied the nonlinear sigma models on Ricci-flat \kahler
manifolds. These models have vanishing $\beta$-function up to the fourth
order in the perturbation theory~\cite{HullAG,Grisaru}. Despite the
appearance of non-zero
$\beta$-function at the four-loop order, we will be able
to obtain the conformally invariant field theory for the background
metric related to the Ricci-flat manifolds through non-local field
redefinition~\cite{Sen}.

After the completion of this work we came to know
that our metric was also discussed in other context~\cite{St,CGLP}. 
Let us give a comment on the relation to their work. 
Defining the new variable $u$ by 
the relation $(a^2 \sinh 2 u)^2 = t^2 - a^4$ in Eq.~(\ref{sol-N}), 
the integral can be rewritten as  
$2 a^{2(N-2)} \int_0^r (\sinh 2u)^{N-2} du 
= 2 a^{2(N-2)} R(r)$, where $r$ has been defined by 
$(a^2 \sinh 2 r)^2 = x^2 - a^4$. 
Here $r$ and $R(r)$ are the same ones introduced in \cite{CGLP}.
We obtain the \kahler potential differentiated by $r$, given by
\begin{align}
\Td{{\cal K}}{r} \ &= \ 2 a^2 \Big\{ \frac{2c}{a^2} (N-1)
\Big\}^{\frac{1}{N-1}} R(r)^{\frac{1}{N-1}} \; .
\end{align} 


\section*{Acknowledgements} 

We would like to thank Gary Gibbons 
for pointing out references~\cite{CGLP}. 
We are also grateful to Takashi Yokono 
for useful comments.


\end{document}